\documentclass[12pt,preprint]{aastex}

\newcommand{\kms}{{{\rm \,km~s^{-1}}}}

\newcommand{\beq}{\begin{equation}}              
\newcommand{\eeq}{\end{equation}}                
\newcommand{\Om}{\Omega_{\rm m,0}}
\newcommand{\Ox}{\Omega_{x,0}}

\shorttitle
{Cosmological Parameters From Strong Lensing}

\shortauthors{K. Chae}

\begin{document}

\title{Cosmological Parameters from the SDSS DR5 Velocity Dispersion Function 
 of Early-type Galaxies through Radio-Selected Lens Statistics}

\author{Kyu-Hyun Chae\altaffilmark{1}}
\altaffiltext{1}{Sejong University, Department of Astronomy and 
Space Sciences, 98 Gunja-dong, Gwangjin-Gu, Seoul 143-747, Republic of Korea;
chae@sejong.ac.kr}

\begin{abstract}

We improve strong lensing constraints on cosmological parameters in light of 
the new measurement of the velocity dispersion function of early-type galaxies
based on the SDSS DR5 data and recent semi-analytical modeling of galaxy
formation. Using both the number statistics of the CLASS 
statistical sample and the image separation distribution of the CLASS and the 
PANELS radio-selected lenses, we find the cosmological matter density 
$\Om = 0.25^{+0.12}_{-0.08}$ (68\% CL) assuming evolutions of galaxies 
predicted by a semi-analytical model of galaxy formation and 
$\Om = 0.26^{+0.12}_{-0.08}$ assuming no evolution of galaxies for a flat 
cosmology with an Einstein cosmological constant. For a flat cosmology with a 
generalized dark energy, we find the non-evolving dark energy equation of state
 $w_x < -1.2$ ($w_x < -0.5$) at the 68\% CL (95\% CL). 
\end{abstract}

\section{Introduction}
Strong lensing has been an important astrophysical tool for probing both
cosmology (e.g., \citealt{Ref64, TOG84, Fuk92, Koc93, Cha02, Cha03, Cha04a, 
Mit05, Yor05}) and galaxies (their structures, formations, and evolutions; 
e.g., 
\citealt{KKS97,MS98,Kee01,KW01,CM03,Ofe03,RK05,Cha05,Tre06,Koo06,Cha06}). 
Strong lensing is also potentially a useful tool to test theories of gravity 
(e.g., \citealt{KP05,KP06}). 

At the time of this writing there are $\sim 90$ galactic-scale strong 
lenses.\footnote{http://cfa-www.harvard.edu/castles}
Parts of them form well-defined samples that are useful for statistical 
analyses. For example, 26 lenses from the Cosmic Lens ALL-Sky Survey 
(CLASS; \citealt{Mye03,Bro03}) and the PMN-NVSS Extragalactic Lens Survey 
(PANELS; \citealt{Win01}) form a well-defined radio-selected lens sample, 
and the Sloan Digital Sky Survey (SDSS; \citealt{Ogu06}) has accumulated 25 
galactic-scale lenses (including 8 re-discoveries) so far and expect to 
eventually obtain 
a statistical sample of from 60-70 lenses.\footnote{N. Inada, 2006 SDSS 
collaboration meeting in Seoul (\mbox{http://astro.snu.ac.kr/~sdss/}).}
 These well-defined samples are particularly useful not only for 
constraining cosmological parameters such as the present-day matter density 
$\Om$, dark energy density $\Ox$ and its equation of state $w_x$ 
(e.g., \citealt{Cha02, Cha03, Mit05}) 
but also for constraining the statistical properties of galaxies such as  
optical region velocity dispersions (e.g., \citealt{Cha05,Cha06}) and galaxy 
evolutions (e.g., \citealt{CM03,Ofe03}).

The sample from the completed CLASS, in particular its subsample of 13 lenses
strictly satisfying well-defined selection criteria (the CLASS statistical 
sample; \citealt{Bro03,Cha03}), was first extensively analyzed by 
\citet{Cha02} and \citet{Cha03}, who found $\Om \approx 0.3$ assuming 
a flat cosmology and adopting non-evolving galaxy populations. 
\citet{Mit05} re-analyzed the CLASS statistical sample based on the velocity 
dispersion function (VDF) of early-type galaxies directly derived from the 
SDSS Data Release 1 (DR1; \citealt{Sto02}) galaxies 
(\citealt{She03}). However, \citet{Cha05} finds that the 
\citet{She03} VDF of early-type galaxies would imply a significantly 
underestimated abundance of early-type galaxies based on the Wilkinson 
Microwave Anisotropy Probe (WMAP) 1st year cosmology (\citealt{Spe03}) and
the CLASS statistical sample. Just recently, \citet{Cho06} have made a new 
measurement of the VDF of early-type galaxies based on the much larger SDSS 
Data Release 5 (DR5; \citealt{Ade07})\footnote{The actually used data set is 
called DR4plus which is very similar to the DR5.} galaxies employing a new 
and more reliable method of classifying galaxies (\citealt{PC05}). 
The \citet{Cho06} VDF has a much higher comoving number density of 
early-type galaxies and a different shape for the lower velocity part compared 
with the \citet{She03} VDF. The \citet{Cho06} early-type number density is
in favor of the \citet{Cha05} results. 

The goal of this work is to improve strong lensing statistics using the 
SDSS DR5 VDF of early-type galaxies. Our focus shall be to put independent
constraints on $\Om$ and $w_x$ assuming a flat cosmology. We shall consider 
both no evolution and a evolution of galaxies based on the prediction by a 
semi-analytical model of galaxy formation (\citealt{Kan05,Cha06}). 
In \S 2, we briefly describe the data and the analysis method. We present 
and discuss the results in \S 3.

\section{Data and Method}

The comoving number density of galaxies as a function of velocity dispersion 
($\sigma$) can be described by the modified Schechter function $\phi(\sigma)$ 
given by (\citealt{She03, Mit05})
\beq
dn = \phi(\sigma) d\sigma = \phi_*
  \left(\frac{\sigma}{\sigma_*}\right)^{\alpha}
   \exp\left[-\left(\frac{\sigma}{\sigma_*}\right)^{\beta}\right]
   \frac{\beta}{\Gamma(\alpha/\beta)} \frac{d\sigma}{\sigma},
\label{VDF}
\eeq
where $\phi_*$ is the integrated number density of galaxies,
 $\sigma_*$ is the
characteristic velocity dispersion, $\alpha$ is the low-velocity power-law
index, and $\beta$ is the high-velocity exponential cut-off index. 
\citet{She03} (the number density being updated by \citealt{Mit05}) found from
the SDSS DR1 for the early-type galaxy population
\begin{eqnarray}
 (\phi_*,\hspace{0.2cm} \sigma_*,\hspace{0.2cm} \alpha,
 \hspace{0.2cm} \beta)_{\rm DR1}
  & = & [(4.1 \pm 0.3) \times 10^{-3} \hspace{0.2cm}h^3
  \hspace{0.2cm}{\rm Mpc}^{-3}, \nonumber  \\
  &   & 88.8 \pm 17.7 \hspace{0.2cm} \kms, \nonumber \\
  &   & 6.5 \pm 1.0, \hspace{0.2cm} 1.93 \pm 0.22],
\label{DR1}
\end{eqnarray}
where $h$ is the Hubble constant in units of 100~$\kms$~Mpc$^{-3}$. This was
the first direct measurement of the VDF of early-type galaxies. Just recently,
\citet{Cho06} have found from the much larger SDSS DR5
\begin{eqnarray}
 (\phi_*,\hspace{0.2cm} \sigma_*,\hspace{0.2cm} \alpha,
 \hspace{0.2cm} \beta)_{\rm DR5}
  & = & [8.0 \times 10^{-3} \hspace{0.2cm}h^3
  \hspace{0.2cm}{\rm Mpc}^{-3}, \nonumber  \\
  &   & 161 \pm 5 \hspace{0.2cm} \kms, \nonumber \\
  &   & 2.32 \pm 0.10, \hspace{0.2cm} 2.67 \pm 0.07].
\label{DR5}
\end{eqnarray}
The above VDFs are intrinsic functions derived taking into account the 
correlated measurement errors of $\sigma$ (\citealt{She03,Cho06}). 
The DR1 and DR5 VDFs are shown in Figure~\ref{fig:Om}(a).
The DR5 VDF is clearly quite different from the DR1 VDF both in the number
density and the shape for the lower part of $\sigma$. 
This is in large part due to the improved galaxy 
classification scheme of \citet{PC05}, who makes use of a SDSS $u$-$r$ color
versus $g$-$i$ color gradient space.

While early-type galaxies dominate strong lensing, late-type galaxies cannot
be neglected. Among the radio-selected lenses with known galaxy types from the 
CLASS and the PANELS (see Table~1 of \citealt{Cha05}), 
about 30\% are late-types. For the late-type galaxy population, 
the direct measurement of the VDF is complicated by the significant rotations 
of the disks. We proceed as follows. We adopt the luminosity function (LF) of 
late-type galaxies from the SDSS DR5 (\citealt{Cho06}). We turn the LF into a 
circular velocity function using a Tully-Fisher relation (\citealt{TP00}). 
Finally, we turn the circular velocity function into a VDF assuming that the 
circular velocity is proportional to the inner velocity dispersion as would be
the case in an isothermal model. In principle, we can estimate all the 
parameters of equation~(\ref{VDF}) for the late-type population
using the Tully-Fisher relation and a galaxy model. However, we shall 
leave $\sigma_*$ free and determine it from the image separation distribution. 
Our adopted VDF for the late-type galaxy population is then
\begin{eqnarray}
 (\phi_*,\hspace{0.2cm} \sigma_*,\hspace{0.2cm} \alpha,
 \hspace{0.2cm} \beta)_{\rm late}
  & = & [1.13 \times 10^{-1} \hspace{0.2cm}h^3
  \hspace{0.2cm}{\rm Mpc}^{-3}, \nonumber  \\
  &   & \sigma_*^{\rm(late)}, \hspace{0.2cm} 0.3, \hspace{0.2cm} 2.91].
\label{late}
\end{eqnarray}
A late-type VDF with $\sigma_*^{\rm(late)}=133\kms$ determined from lensing in 
\S 3 is shown in Figure~\ref{fig:Om}(a).
It is interesting to notice that this VDF of late-type galaxies matches 
relatively well that of \citet{She03} who determined $\sigma_*^{\rm(late)}$ 
using a Tully-Fisher relation taking into account the scatter of the 
Tully-Fisher relation.

The SDSS galaxy populations refer to a redshift range of $0 \la z \la 0.2$ 
while the radio-selected lenses are in the range of $0.3 \la z \la 1$, so that 
galaxy evolutions must be taken into account. Most of previous works have 
been done with the assumption of no evolution of early-type galaxies from 
$z=0$-1 relying on several observational arguments (see \citealt{Cha03} and 
references therein) such as the fundamental plane and the color-magnitude 
relations. More recently, \citet{Mit05} have taken into consideration galaxy 
evolutions using the prediction by the extended Press-Schechter model of 
structure formation which is calibrated by N-body simulations (\citealt{ST02}).
We take into consideration galaxy evolutions using the predictions by a recent 
semi-analytical model of galaxy formation (\citealt{Kan05,Cha06}) which is 
based on the high-resolution N-body simulations of \citet{JS02}. Specifically,
we constrain the evolutions of the VDFs using the evolutions of the virial 
circular velocity functions used by \citet{Cha06} assuming a power-law relation
between the virial circular velocity and the velocity dispersion as given
by equation~(6) of \citet{Cha06}. We consider the evolutions of both the number
density $\phi_*$ and the characteristic velocity dispersion $\sigma_*$ as 
follows:
\beq
 \phi_*(z)=\phi_{*,0} (1+z)^{\nu_n};\hspace{0.2cm} 
 \sigma_*(z)=\sigma_{*,0} (1+z)^{\nu_v}.
\label{evmod}
\eeq
We obtain the following best-fit parameters for the early-type and the 
late-type galaxy populations of \citealt{Kan05}:
\beq
 (\nu_n,\hspace{0.2cm} \nu_v) = (-0.229, -0.01)\hspace{0.2cm} 
 \mbox{for early-type and} \hspace{0.2cm}  (1.24, -0.186)\hspace{0.2cm} 
  \mbox{for late-type}.
\label{evparam}
\eeq

We use all the lensing properties of the CLASS and the PANELS lenses 
(Table~1 of \citealt{Cha05}). Specifically, we use both the image separation 
distribution of all `single' lenses as well as the number statistics of the
well-defined CLASS statistical sample of 13 lenses. We use the same singular
isothermal ellipsoid lens model and analysis method recently used by 
\citet{Cha03}, \citet{Cha05}, and \citet{Cha06}. The likelihood function is 
essentially the same as that given by equation~(11) of \citet{Cha06}, namely
 \beq
\ln \mathcal{L} =
          \left( \sum_j \ln \delta p_{\rm IS}(j) \right)
                + \left( \sum_k \ln [1 - p(k)]
                + \sum_l \ln \delta p(l) \right),
\label{Lhood}
\eeq
where $\delta p_{\rm IS}(j)$ is the relative image separation probability, 
$p(k)$ is the total multiple-imaging probability due to both the early-type and
the late-type populations, and $\delta p(l)$ is the differential lensing 
probability of the specific separation, lensing galaxy type, lens and source 
redshifts. The only difference here is that for the lenses of unknown galaxy 
morphologies the image separation probability $\delta p_{\rm IS}(j)$ is 
calculated as a weighted sum due to the early-type and the late-type galaxy 
populations where the weighting factors are the differential lensing 
probabilities for the observed specific image separations 
({eq.}~29 of \citealt{Cha03}).

\section{Results and Discussion}

\begin{figure}
\epsscale{0.65}
\plotone{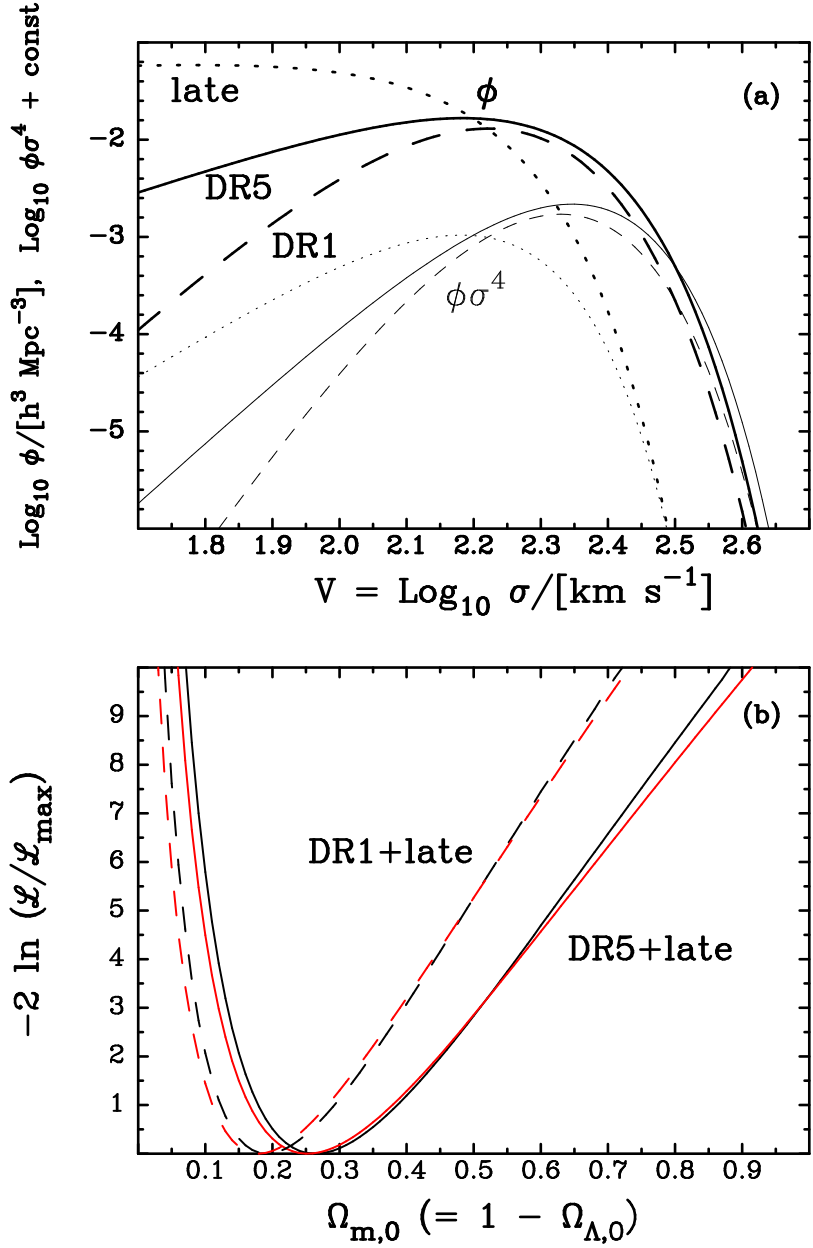}
\caption{(a) Thick lines are the DR5 (solid) and the DR1 (dashed) early-type
 VDFs and the late-type VDF (dotted). Thin lines show the behavior of 
$\phi \sigma^4$ (which is proportional to the differential lensing probability)
for these VDFs. 
(b) The behavior of the likelihood function (eq.~\ref{Lhood}) as $\Om$ is 
varied for a flat cosmology with an Einstein cosmological constant.
The solid (dashed) lines are based on the SDSS DR5 (DR1) VDF of early-type
galaxies and the SDSS DR5 LF of late-type galaxies. The black and red curves
correspond respectively to the cases of assuming no evolution of galaxies
and the evolutions of galaxies predicted by a recent semi-analytical model of
galaxy formation. 
}
\label{fig:Om}
\end{figure}

We first consider a flat cosmology with an Einstein cosmological constant.
Figure~\ref{fig:Om}(b) shows the behavior of the likelihood function 
(eq.~\ref{Lhood}) as the matter density $\Om$ is varied. The results shown
in Figure~\ref{fig:Om}(b) are based on the SDSS DR1/DR5 early-type VDF 
(eqs.~\ref{DR1} \& \ref{DR5}) and the late-type VDF (eq.~\ref{late}). 
For the DR5 VDF we find $\Om=0.25^{+0.12}_{-0.08}$ and
 $0.26^{+0.12}_{-0.08}$ (68\% CL) respectively for the evolving 
(eq.~\ref{evparam}) and non-evolving populations of galaxies. 
The fitted value of $\sigma_{*,0}^{\rm(late)}$ for the late-type population is 
133~$\kms$ and 127~$\kms$ respectively for the evolving and non-evolving cases.

For the DR1 VDF we find $\Om=0.18^{+0.1}_{-0.06}$ and
 $0.19^{+0.1}_{-0.06}$ (68\% CL) respectively for the evolving 
(eq.~\ref{evparam}) and non-evolving populations of galaxies.
The difference between the DR1 and the DR5 results can be understood from
the behavior of $\phi \sigma^4$, which is proportional to the differential
lensing probability, shown in Figure~\ref{fig:Om}(a).
Notice that our results based on the DR1 VDF are different from the 
\citet{Mit05} results because of differences in the calculations of 
magnification biases and cross sections satisfying observational selection
functions. For example, for the differential number-flux density relation of 
$|dN/dS| \propto (S/S_0)^{-\eta}$ with $S_0=30$~mJy, we use $\eta=1.97$ (2.07)
for $S < S_0$ ($S > S_0$) while \citet{Mit05} uses erroneously $\eta=2.1$
 for any $S$.\footnote{While the lensed sources have observed flux densities 
$S_{\rm ob} > 30$~mJy, the magnification factor is given as an integration
for the range from $S = S_{\rm ob}$ to $S_{\rm ob}/\mu_{\rm max}$ where 
$\mu_{\rm max}$ is the maximum possible theoretical magnification (see eq.~36 
of \citealt{Cha03}). Thus, it is essential to use $\eta=1.97$ for 
$S < 30$~mJy.} Taking into account the flux ratio limit for the doubly-imaged
systems in the CLASS statistical sample, namely that the fainter-to-brighter 
image flux ratio must be greater than 0.1 (\citealt{Bro03,Cha03}), 
\citet{Mit05} finds  $\tilde{B} = 3.97$ using the singular isothermal sphere 
(SIS) model, where $\tilde{B}$ (eq.\ 15 of \citealt{Mit05}) is the
magnification bias times the cross section satisfying the flux ratio limit
 divided by the unbiased cross section. However, we find 
$\tilde{B} \approx 3.36$ for the SIS taking correctly into account the 
CLASS observational selection functions.  Another difference is that 
\citet{Mit05} uses the SIS while we use the singular isothermal ellipsoid 
(SIE). For example, a lens ellipticity of 0.4 can amount to a difference of 
$\Delta \Om \approx -0.05$ compared with the spherical case because of the 
variation of the magnification bias and cross section for the equal numbers of
oblates and prolates (see, however, \citealt{Hut05} for other possibilities).

The directly 
measured SDSS DR5 VDF of early-type galaxies is more reliable than VDFs 
inferred from early-type LFs using a Faber-Jackson (\citealt{FJ76}) relation 
(e.g., \citealt{Cha03}) because of significant scatters in the relation 
(\citealt{She03}).  The SDSS DR5 VDF is also likely to be more reliable than
the DR1 VDF (\citealt{She03}) not only because the DR5 VDF is based on a much
larger volume sample but also because it is based on a more reliable galaxy
classification technique by \citet{PC05}. Therefore, the SDSS DR5 VDF 
in conjunction with galaxy evolution models from a recent semi-analytic
model of galaxy formation (\citealt{Kan05, Cha06}) removes in large part 
potential systematic errors in one of the main ingredients of strong lensing 
statistics.

Our derived values of the matter density based on the SDSS DR5 VDF 
are in excellent agreement with the results from 
the WMAP and the large-scale structures in the SDSS luminous red galaxies 
(\citealt{Spe03,Spe06,Teg04,Eis05}).
The SDSS lens search is eventually expected to discover from 60-70 strongly 
lensed quasars. Thus, we expect that the precision of strong lensing 
statistics will improve by a factor of two or better in the near future 
(in particular if the CLASS statistical sample is combined with the SDSS 
sample). Furthermore, within a few decades next generation observation tools
such as the Square Kilometre Array (e.g., \citealt{Bla04,Koo04}) will improve
the precision of lensing statistics by orders of magnitude. 
Perhaps, strong lensing statistics may play even more important roles 
in the future for uncovering the physical processes of galaxy formations and 
evolutions (see, e.g., \citealt{KW01,Kee01,CM03,Ofe03,Cha06}).

\begin{figure}
\epsscale{0.8}
\plotone{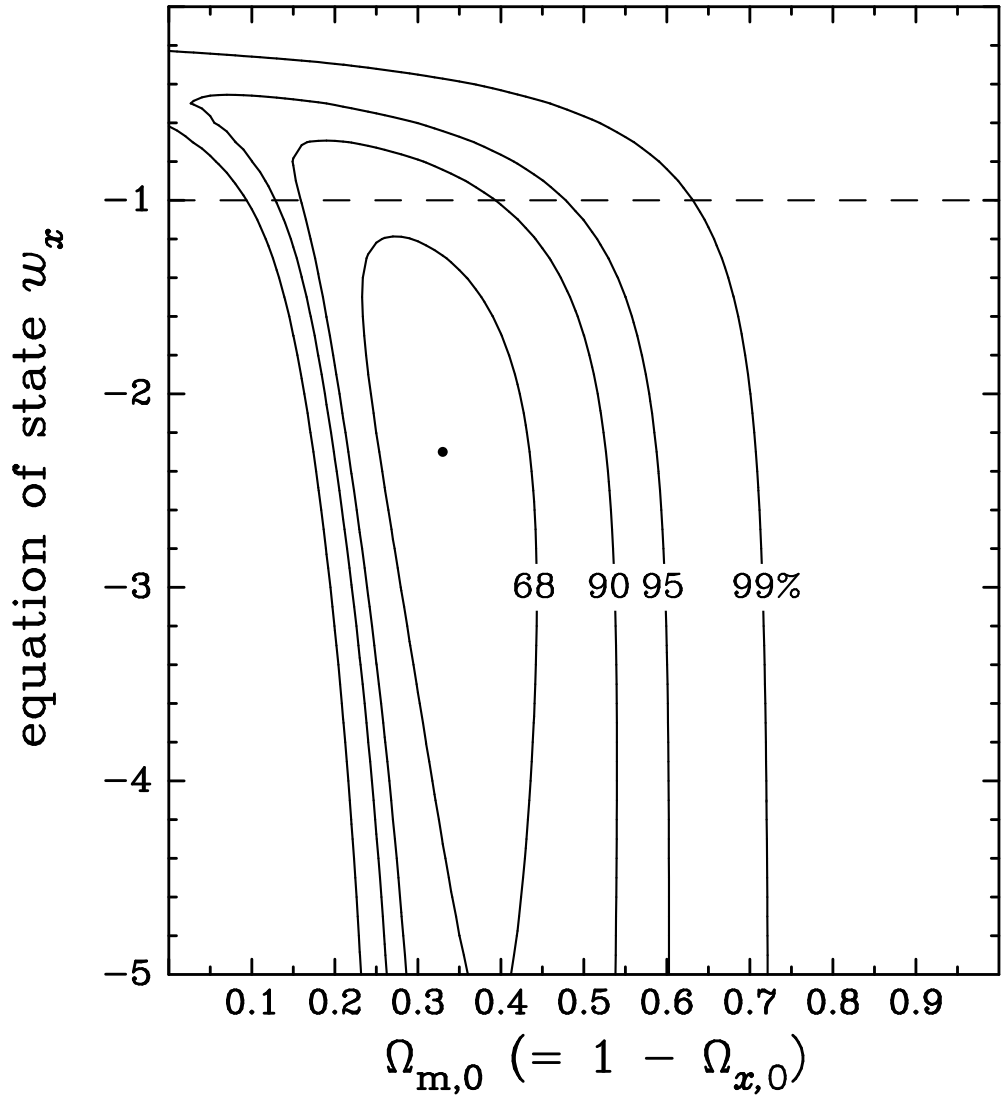}
\caption{
Constraints in the plane spanned by the matter density $\Om$ ($= 1 - \Ox$ in 
our assumed flat cosmology) and the non-evolving equation of state $w_x$ of
a generalized dark energy based on the SDSS DR5 VDF of early-type galaxies and 
LF of late-type galaxies through radio-selected strong lensing statistics.
We have assumed the evolutions of galaxies predicted by a semi-analytical model
of galaxy formation.
}
\label{fig:Omw}
\end{figure}

Finally we consider a flat cosmology with a generalized dark energy $\Ox$ with 
a non-evolving equation of state $w_x$. Figure~\ref{fig:Omw} shows the
confidence limits in the $\Om$-$w_x$ plane. We have $w_x < -1.2$ at the 68\% CL
($w_x < -0.5$ at the 95\% CL), so that strong lensing appears to marginally 
favor a super-negative equation of state (e.g., \citealt{Cal03}). 
 On the other hand, the WMAP results (\citealt{Spe06}) favor 
 $w_x = -1$ (Einstein's cosmological constant).  
It is not clear at the present why lensing data appear
to favor $w_x < -1$. It will be interesting to see whether this remains 
so with future lensing data.

\bigskip

K.-H. C. thanks C. Park, Y.-Y. Choi and C.-G. Park for useful discussions about
the SDSS DR5 velocity dispersion function of early-type galaxies. The author
also thanks the anonymous referee for pointing out the correlated 
measurement errors of velocity dispersions. 
This work is supported by the Astrophysical Research Center for the Structure 
and Evolution of the Cosmos (ARCSEC). 

Funding for the SDSS and SDSS-II has been provided by the Alfred 
P. Sloan Foundation, the Participating Institutions, the National Science 
Foundation, the U.S. Department of Energy, the National Aeronautics and Space 
Administration, the Japanese Monbukagakusho, the Max Planck Society, and the 
Higher Education Funding Council for England. The SDSS Web Site is 
http://www.sdss.org/. 

The SDSS is managed by the Astrophysical Research 
Consortium for the Participating Institutions. The Participating Institutions 
are the American Museum of Natural History, Astrophysical Institute Potsdam, 
University of Basel, University of Cambridge, Case Western Reserve University, 
University of Chicago, Drexel University, Fermilab, the Institute for Advanced 
Study, the Japan Participation Group, Johns Hopkins University, the Joint 
Institute for Nuclear Astrophysics, the Kavli Institute for Particle 
Astrophysics and Cosmology, the Korean Scientist Group, the Chinese Academy of 
Sciences (LAMOST), Los Alamos National Laboratory, the Max-Planck-Institute for
 Astronomy (MPIA), the Max-Planck-Institute for Astrophysics (MPA), New Mexico 
State University, Ohio State University, University of Pittsburgh, University 
of Portsmouth, Princeton University, the United States Naval Observatory, and 
the University of Washington.

\clearpage

\clearpage

\end{document}